\documentclass[preprint,showpacs,titlepage,amsmath,amssymb]{revtex4-2}
\usepackage{graphicx}
\usepackage{dcolumn}
\usepackage{bm}
\usepackage{hyperref}

\begin{document}

\title{The yield ratio of anti-nuclei and nuclei of relativistic nuclear collisions in the central rapidity region}

\author{A.I. Malakhov}
\email{malakhov@jinr.ru}
\affiliation{Veksler and Baldin Laboratory of High Energy Physics Joint Institute for Nuclear Research, Dubna, Russia}

\author{A.A. Zaitsev}
\affiliation{Veksler and Baldin Laboratory of High Energy Physics Joint Institute for Nuclear Research, Dubna, Russia}


\begin{abstract}
This article describes calculations of the yield ratio of anti-particles to the yield of particles ($\bar{p}/p$, $\bar{d}/d$,$\overline{^{3}\mathrm{He}}$/$^{3}$He) in proton-proton and nuclear-nuclear interactions using the self-similarity parameter in the central rapidity region. The used approach is based on the study of relativistic nuclear interactions in the four-velocity space. The results of the calculations are compared with the existing experimental data in a wide center-of-mass energy range ($SPS$, $RHIC$, $LHC$). Within this approach the inclusive spectra of pions and kaons and ratios of their yields in $pp$ collisions have been successfully described earlier. 
\end{abstract}

\maketitle 

\section*{Introduction}
\noindent 
In 1998 A.M. Baldin and A.I. Malakhov published an article where the self-similarity was applied to obtain an explicit analytical expression for inclusive cross sections of particles, nuclear fragments and anti-nuclei in relativistic nuclear collisions in the central rapidity region \cite{BaldinMalakhov}. At present the corresponding experimental data have appeared up to the LHC energy. In this article we give a comparison of our calculations concerning the yield ratio of anti-nuclei and nuclei in $pp$ and central $AA$ relativistic collisions in the mid-rapidity region with the available experimental data.

\section*{Self-similarity parameter in the central rapidity region}
The experimentally observable quantities which characterize multi-quark processes are the cross sections of multiple particle production in relativistic nuclear collisions:

\begin{equation} 
\label{eq:1}
I  +  II   \to  1 + 2 + 3 + …
\end{equation}

In reaction (1), the colliding nuclei are depicted with numbers I and II, and secondary particles are indicated with numbers 1, 2, 3... . The process (1) is schematically presented in Fig.1. 

\begin{figure}[h]
	\centerline{\includegraphics*[width=0.4\linewidth]{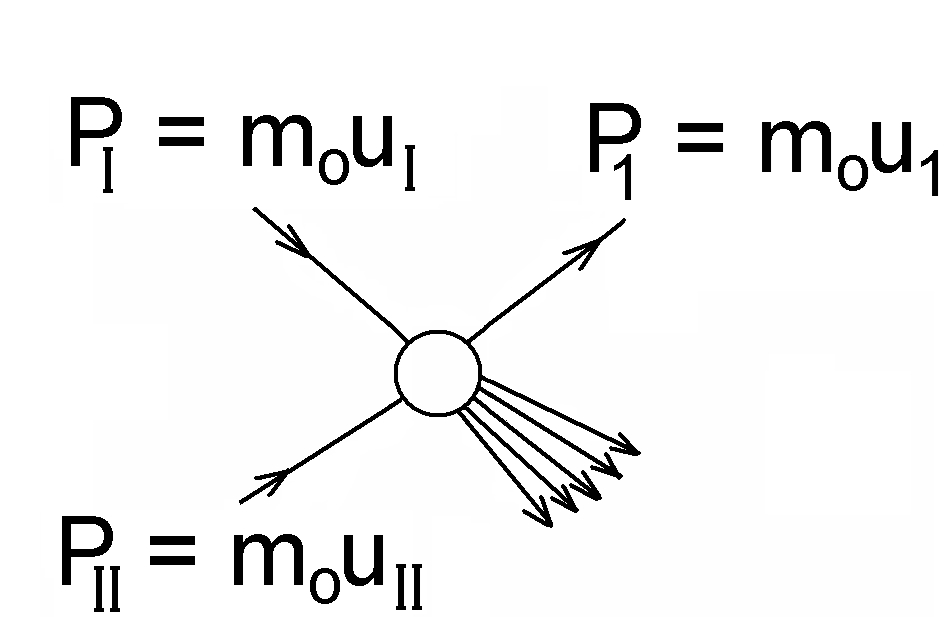}}
	\caption{Interaction of two nuclei: P$_{\textrm{I}}$ is 4-momentum of nucleus I; P$_{\textrm{II}}$ is 4-momentum of nucleus II, and P$_{1}$ is 4-momentum of secondary particle 1; u$_{\textrm{I}}$, u$_{\textrm{II}}$, u$_{1}$ are 4-velocities of nuclei I, II and particle 1; $m_0$ is  a nucleon mass.}
	\label{fig:Fig.1}
\end{figure}

For the process (1), if we observe only one secondary particle, it is possible to write the conservation law of 4-momentum in the following form:

\begin{equation} 
\label{eq:2}
(N_\textrm{I}\cdot P_\textrm{I} + N_{\textrm{II}}\cdot P_{\textrm{II}} – P_1)^2 = (N_\textrm{I}\cdot m_0 + N_{\textrm{II}}\cdot m_0 + M)^2
\end{equation}

Here $M$ is the mass of the particle providing conservation of the baryon number, strangeness, and other quantum numbers. For $\pi$ mesons $m_1$ = $m_{\pi}$ and $M$ = 0. For $K^-$ mesons $m_1$ = $m_K$  and $M$ = $m_K$. For $K^+$ mesons $m_1$ = $m_K$ and $M$ = $m_\Lambda$ – $m_K$, $m_\Lambda$ is the mass of $\Lambda$ baryon. 

To describe the interactions of relativistic nuclei in the four-velocity space, the self-similarity parameter was introduced into \cite{BaldinBaldin}:

\begin{equation} 
\label{eq:3}
\Pi = min\left\lbrace \frac{1}{2} \sqrt{\left( u_\textrm{I}\cdot N_\textrm{I}  + u_{\textrm{II}}\cdot N_{\textrm{II}}\right)^2} \right\rbrace ,  
\end{equation}

\noindent where $N_\textrm{I}$ and $N_{\textrm{II}}$ are cumulative numbers (or more precisely, the fractions of the transferred 4-momentum) for nuclei I and II, and $u_\textrm{I}$ and $u_{\textrm{II}}$ are 4-velocities of these nuclei.

In this case the invariant cross-sections of the output inclusive particles of different types at nuclear interactions with atomic numbers $A_\textrm{I}$ and $A_{\textrm{II}}$, are described by universal dependence in a broad energy range and different atomic numbers of the colliding nuclei \cite{BaldinBaldin,BaldinMalakhovSissakian}:

\begin{equation} 
\label{eq:4}
E\cdot d^3\sigma/dp^3 = C_1\cdot A_\textrm{I}^{\alpha(N_\textrm{I})} \cdot A_{\textrm{II}}^{\alpha(N_{\textrm{II}})} \cdot \textrm{exp}(-\Pi/C_2) ,
\end{equation}

\noindent where $\alpha(N_\textrm{I})$ = 1/3 + $N_\textrm{I}$/3, $\alpha(N_{\textrm{II}})$ = 1/3 + $N_{\textrm{II}}$/3,\\
$C_1$ = 1.9$\cdot$10$^4$ mb$\cdot$GeV$^{-2}\cdot$c$^3$$\cdot$st$^{-1}$ and $C_2$ = 0.125$\pm$0.002.

Since $E·d^3\sigma/dp^3$ = $d^3\sigma/(d\phi\cdot dy\cdot p_t\cdot dp_t)$, integrating expression (4) by azimuth angle $\phi$, we obtain the following:

\begin{equation} 
\label{eq:5}
d^2\sigma/(m_t\cdot dm_t\cdot dy) = 2\pi\cdot C_1\cdot A_\textrm{I}^{\alpha(N_\textrm{I})}\cdot A_{\textrm{II}}^{\alpha(N_{\textrm{II}})}\cdot \textrm{exp}(-\Pi/C_2).
\end{equation}

\noindent Then we can write

\begin{equation} 
\label{eq:6}
d^2\sigma/(dm_t\cdot dy) = 2\pi\cdot m_t\cdot C_1\cdot A_\textrm{I}^{\alpha(N_\textrm{I})}\cdot A_{\textrm{II}}^{\alpha(N_{\textrm{II}})}\cdot \textrm{exp}(-\Pi/C_2).
\end{equation}

In the central rapidity region it is possible to find the analytical expression for $\Pi$ \cite{BaldinMalakhov}. For this case $N_\textrm{I}$ and $N_{\textrm{II}}$ are equal to each other $N_\textrm{I}$ = $N_{\textrm{II}}$ = $N$.

\begin{equation} 
\label{eq:7}
N = \left[ 1 + \sqrt{(1 + \Phi_M/\Phi_2)}\right] \cdot \Phi ,
\end{equation}

\noindent where

\begin{equation} 
\label{eq:8}
\Phi = (1/2m_0)\cdot(m_{1\textrm{T}}\cdot \textrm{ch}Y+M)/\textrm{sh}^2Y
\end{equation}
and
\begin{equation} 
\label{eq:9}
\Phi_M = (M^2 – m_1^2)/(4m_0^2\cdot \textrm{sh}^2Y) .
\end{equation}

\noindent Here $m_{1T}$ is transverse mass of particle 1, $m_{1T}$ = $\sqrt{(m_1^2 + p_T^2)}$ \\

\noindent and then we obtain the $Baldin-Malakhov$ $equation$:

\begin{equation} 
\label{eq:10}
\Pi = N\cdot \textrm{ch}Y .
\end{equation}

\section*{The yield ratio of antiparticles to particles}

\noindent From the ratios (7-9) for baryons we obtain the following:

\begin{equation} 
\label{eq:11}
\Pi_b = (m_{1T}\cdot \textrm{ch}Y – m_1)\cdot \textrm{ch}Y/(m_0\cdot \textrm{sh}^2Y)
\end{equation}

\noindent and for anti-baryons:

\begin{equation} 
\label{eq:12}
\Pi_a = (m_{1T}\cdot \textrm{ch}Y + m_1)\cdot \textrm{ch}Y/(m_0\cdot \textrm{sh}^2Y) .
\end{equation}

For the yield ratio of anti-baryons to baryons, we get the following:

\begin{equation} 
\label{eq:13}
\mbox{Ratio}\left( \frac{\mbox{antibaryon}}{\mbox{baryon}}\right) = \frac{2\pi\int\limits_0^\infty m_{1T}\cdot C_{1}\cdot A^{\alpha(N_{\textrm{I}})}_{\textrm{I}}\cdot A^{\alpha(N_{\textrm{II}})}_{\textrm{II}} \cdot \mbox{exp}\left( -\frac{\Pi_{a}}{C_{2}} \right) dm_{1T}}{2\pi\int\limits_0^\infty m_{1T}\cdot C_{1}\cdot A^{\alpha(N_{\textrm{I}})}_{\textrm{I}}\cdot A^{\alpha(N_{\textrm{II}})}_{\textrm{II}} \cdot \mbox{exp}\left( -\frac{\Pi_{b}}{C_{2}} \right) dm_{1T}}
\end{equation}

\noindent In case of symmetric nuclei ($A_\textrm{I}$ = $A_{\textrm{II}}$ = $A$) the above relation takes the following form:

\begin{equation} 
\label{eq:14}
\begin{aligned}
&\textrm{Ratio}\left( \frac{\mbox{antibaryon}}{\mbox{baryon}}\right) = \frac{2\pi\int\limits_0^\infty m_{1T}\cdot C_{1}\cdot A^{2\left(\frac{1}{3} + \frac{1}{3}\left[ \frac{m_{1T}}{m_{0}} chY + \frac{m_{1}}{m_{0}}\right] \right)\frac{1}{sh^{2}Y} } \cdot \mbox{exp}\left( -\frac{\Pi_{a}}{C_{2}} \right) dm_{1T}}{2\pi\int\limits_0^\infty m_{1T}\cdot C_{1}\cdot A^{2\left(\frac{1}{3} + \frac{1}{3}\left[ \frac{m_{1T}}{m_{0}} chY - \frac{m_{1}}{m_{0}}\right] \right)\frac{1}{sh^{2}Y} } \cdot \mbox{exp}\left( -\frac{\Pi_{b}}{C_{2}} \right) dm_{1T}} = \\
& = A^{\frac{4}{3}\frac{m_{1}}{m_{0}}\frac{1}{sh^{2}Y}} \cdot \frac{\int\limits_0^\infty m_{1T}\cdot C_{1}\cdot A^{\frac{2}{3}\frac{m_{1T}}{m_{0}}chY} \cdot \mbox{exp}\left( -\frac{\Pi_{a}}{C_{2}} \right) dm_{1T}}{\int\limits_0^\infty m_{1T}\cdot C_{1}\cdot A^{\frac{2}{3}\frac{m_{1T}}{m_{0}}chY} \cdot \mbox{exp}\left( -\frac{\Pi_{b}}{C_{2}} \right) dm_{1T}} = \\
& = A^{\frac{4}{3}\frac{m_{1}}{m_{0}}\frac{1}{sh^{2}Y}} \cdot \mbox{exp}\left(- \frac{2\frac{m_{1}}{m_{0}}\cdot \frac{chY}{sh^{2}Y}}{C_{2}} \right) .
\end{aligned}.
\end{equation}

\noindent If $A_I$  = $A$,  $A_{II}$ =  $B$, then

\begin{equation} 
\label{eq:15}
\mbox{Ratio}\left(\frac{\mbox{antibaryon}}{\mbox{baryon}}\right) = \left( A\cdot B\right) ^{\frac{2}{3}\frac{m_{1}}{m_{0}}\frac{1}{sh^{2}Y}} \cdot \mbox{exp}\left(- \frac{2}{C_{2}}\frac{m_{1}}{m_{0}}\cdot \frac{chY}{sh^{2}Y} \right).
\end{equation}

\begin{figure}[t]
	\centerline{\includegraphics*[width=0.7\linewidth]{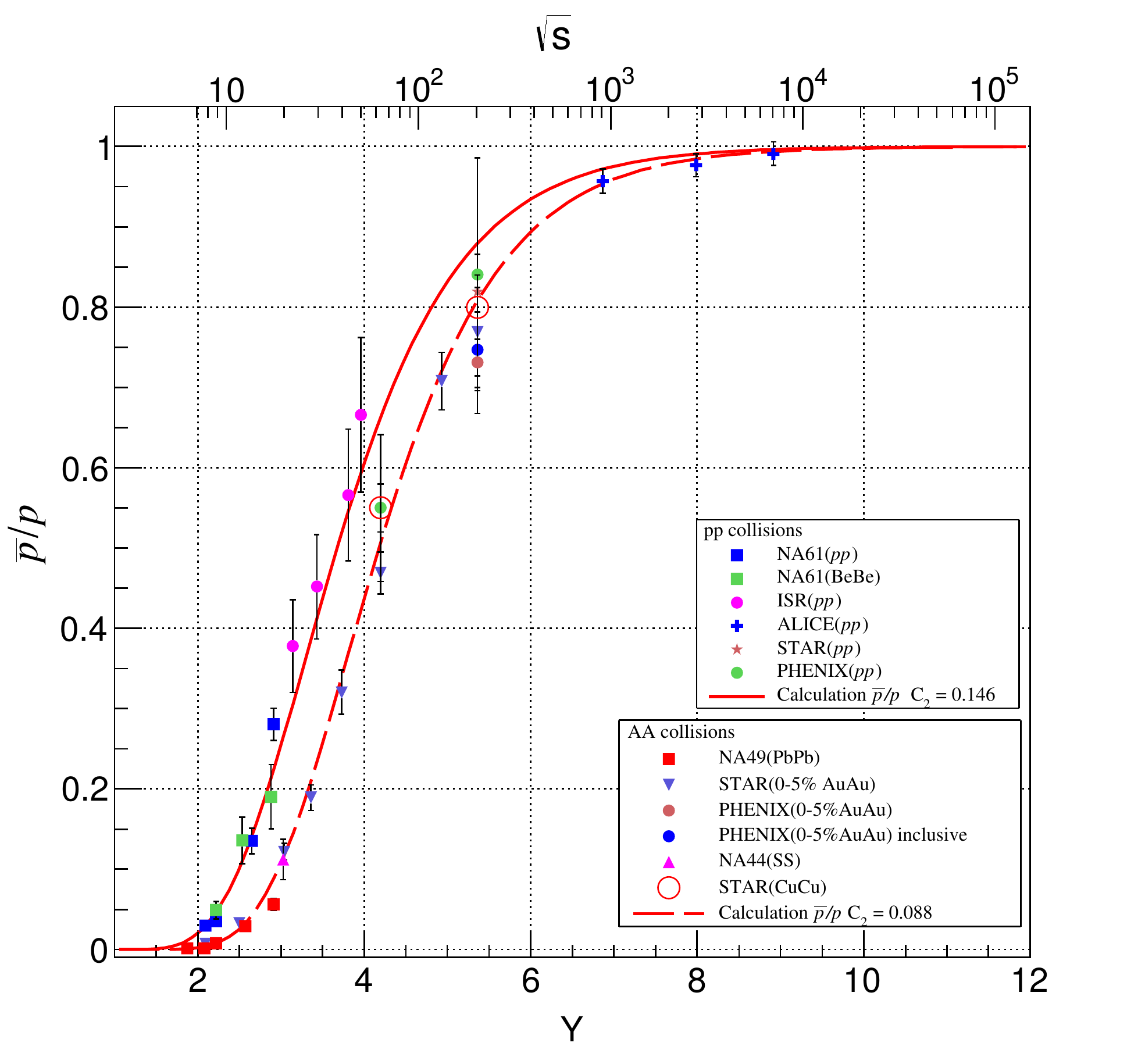}}
	\caption{Dependence of the yield ratios of anti-protons to protons in the central rapidity region on the rapidity and energy in $pp$ \cite{Aduszkiewicz, Rossi, Aamodt, Abelev, Adler, Acharya, Kornasa, Adare, Bearden} and the most central $AA$ \cite{Acharya,Kornasa,Abelev, Adare, Adler,Bearden,Aggarwal} collisions.}
	\label{fig:Fig.2}
\end{figure}

\begin{figure}[t]
	\centerline{\includegraphics*[width=0.95\linewidth]{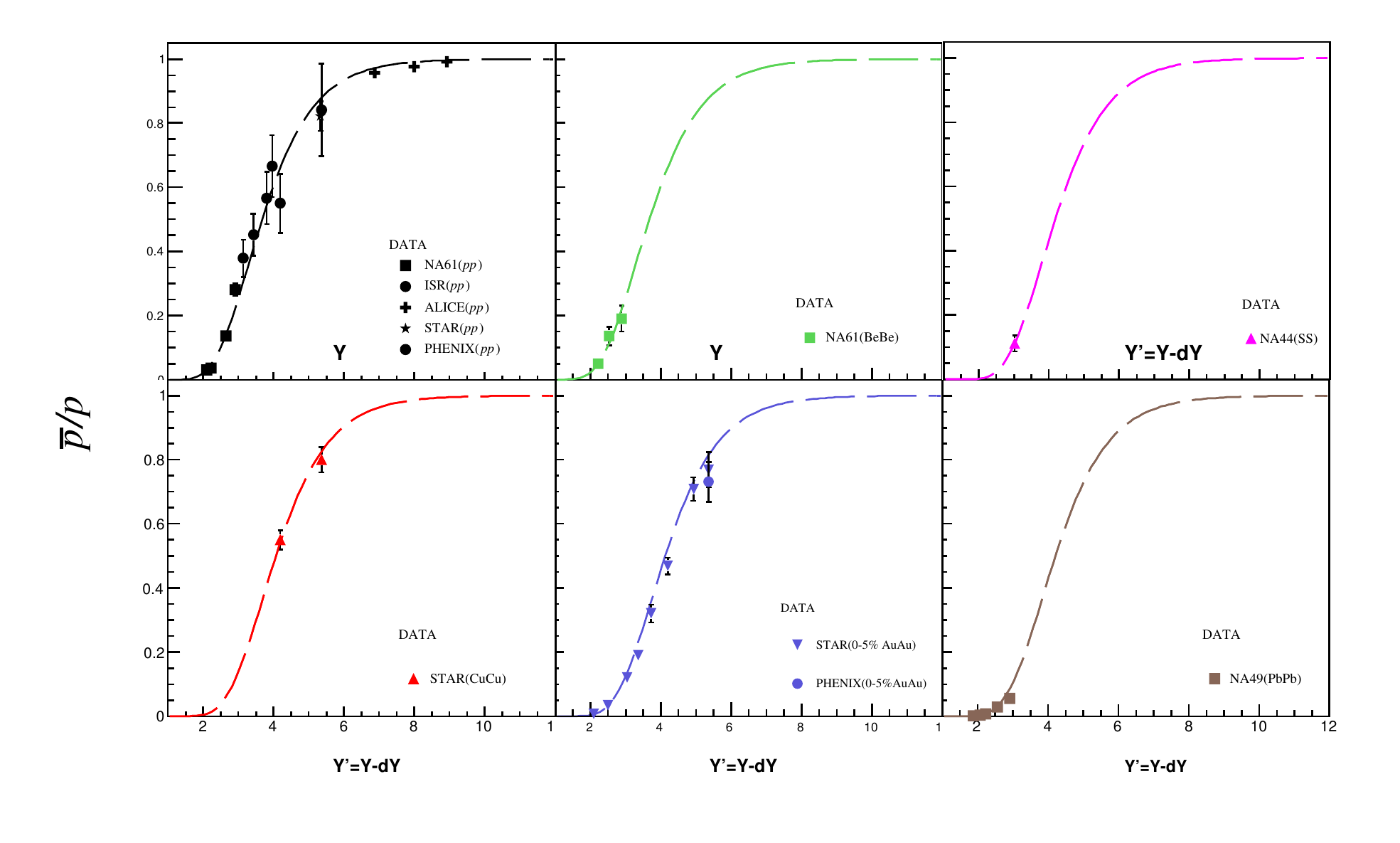}}
	\caption{Dependence of the yield ratios of anti-protons to protons in the central rapidity region on the rapidity of interacting protons of $Y$ for $p+p$ and $Be+Be$ collisions and of $Y'$= $Y$-d$Y$ (d$Y$$\approx$0.5) for $S+S$, $Cu+Cu$, $Au+Au$ and $Pb+Pb$ nuclei.}
	\label{fig:Fig.3}
\end{figure}

\begin{figure}
	\centerline{\includegraphics*[width=0.55\linewidth]{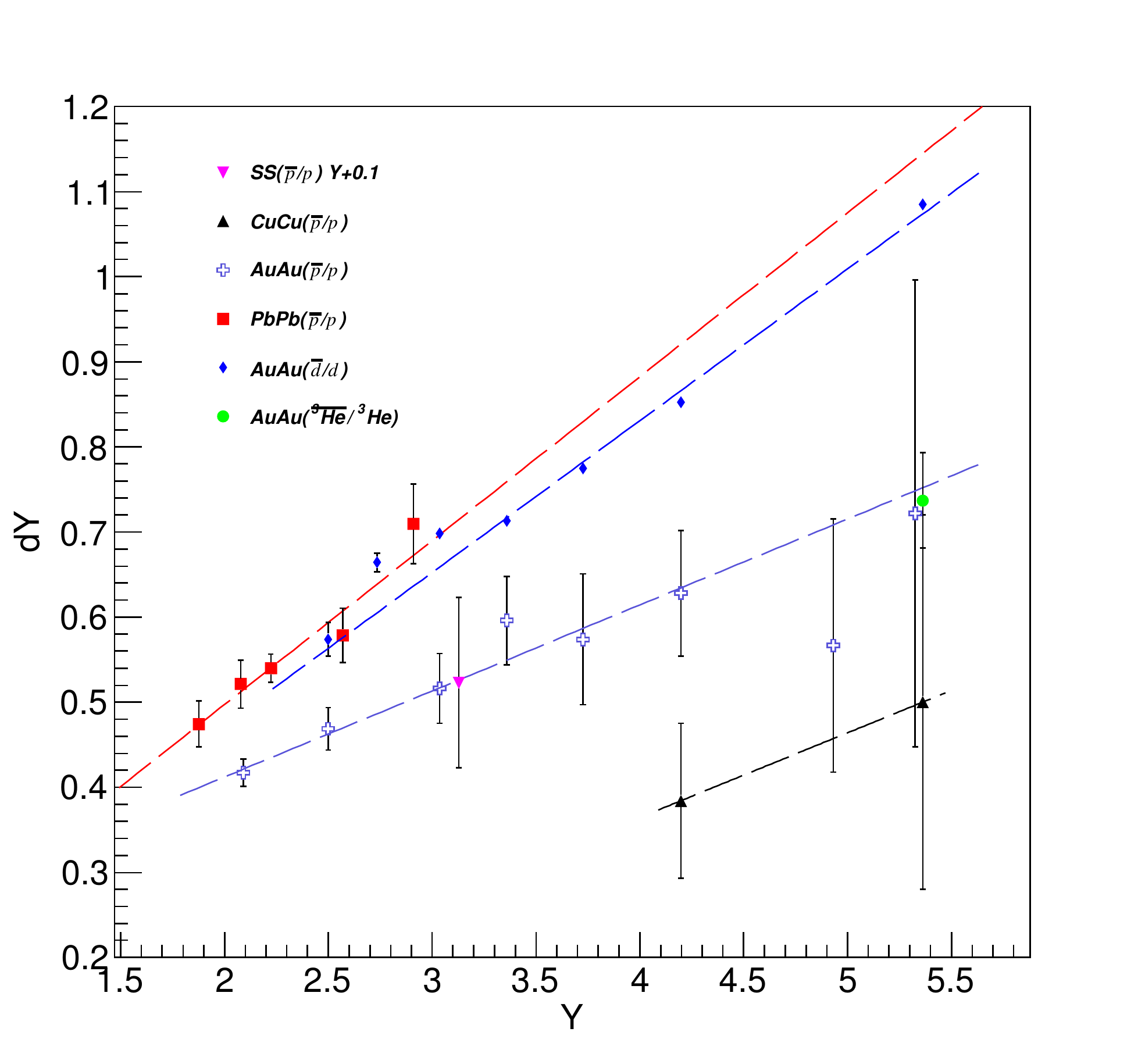}}
	\caption{The dependence of the rapidity loss of d$Y$ on the rapidity of $Y$. The dotted lines - linear approximation of d$Y$($Y$)=$p_0$+$p_1\cdot Y$.}
	\label{fig:Fig.4}
\end{figure}

\begin{figure}[h!]
	\centerline{\includegraphics*[width=1\linewidth]{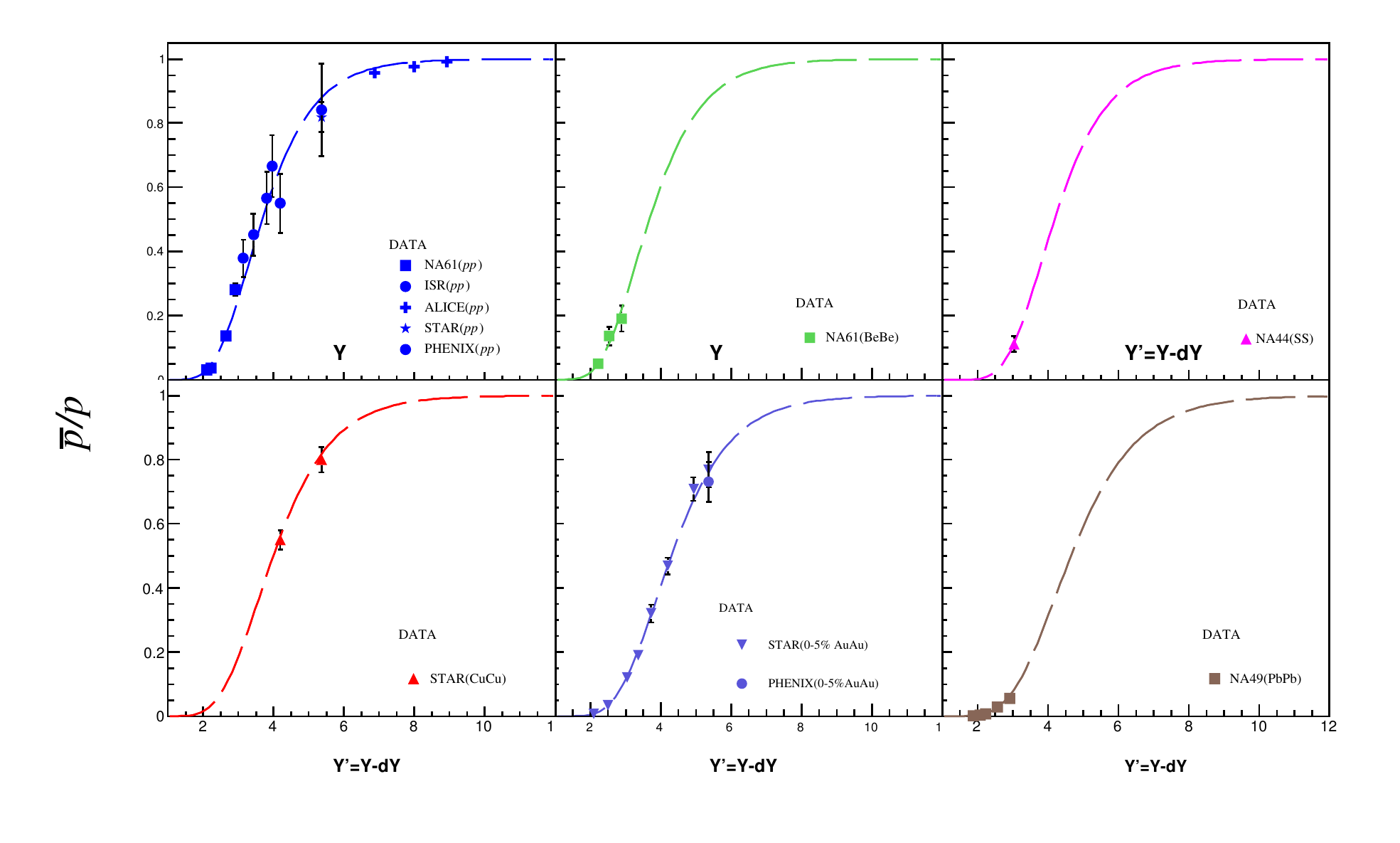}}
	\caption{Description of the yield ratios of $\bar{p}$/$p$ with one value of constant $C_2$=0.146 taking into account d$Y$($Y$) dependences.}
	\label{fig:Fig.5}
\end{figure}

The results of calculations for the yield ratio of anti-protons to protons by using (13) and (11-12) are presented in Fig.2 together with the experimental data including the latest data obtained at the SPS, RHIC, LHC \cite{Aduszkiewicz, Rossi, Aamodt, Abelev, Adler, Acharya, Kornasa, Adare, Bearden, Aggarwal, Adam}. 

The dependences of the ratio of the anti-proton yield to the proton yield in $p$-$p$ and light nuclei ($Be$-$Be$) interactions are well described by means of our approach with the constant $C_2$ = 0.146$\pm$0.003 (Fig.2 solid red line). The obtained value of the constant coincides with the constant from the work on the describing the inclusive spectra of pions and kaons produced in $pp$ collisions at mid-rapidity region \cite{LMZ:2021}.  For the case of nuclear interactions starting with sulfur ($S$-$S$), a satisfactory description has been done with $C_2$ = 0.088$\pm$0.001.

It can be seen that the experimental data on nuclei from $S$+$S$ to $Pb$+$Pb$ are described with a different constant in comparison with proton-proton and $Be$+$Be$ interactions. However, if we take into account the presence of baryon stopping, then it is necessary to introduce a rapidity loss of $Y$ \cite{Weber, Zhou}. After introducing the rapidity loss by approximately the value d$Y$ $\approx$ 0.5, all dependencies for these nuclei are perfectly described using single constant $C_2$~=~0.146 (Fig.3).

In principle, d$Y$ depends on $Y$, so we fitted dependencies for nuclei from $S$+$S$ to $Pb$+$Pb$ for each value of $Y$ and got the dependences shown in Fig.4. Then the points were fitted with a linear approximation of the form: d$Y$=$p_0$+$p_1\cdot Y$.

Fig.5 illustrates experimental data on the yield ratios of anti-nuclei to nuclei using the same constant $C_2$ = 0.146 taking into account d$Y$($Y$) dependences. It is clearly seen that, taking into account the rapidity dependences, the description of the experimental data occurs in the best way for heavy nuclei ($Au+Au$) at low collision energies.

The calculations of dependence of the yield ratios of anti-deuterons to deuterons in the central rapidity region on the rapidity and energy of interacting $Au$+$Au$ nuclei \cite{Adam, dd_ratio_PHENIX} with $C_2$ = 0.146 and constant rapidity shift are shown in Fig.6(left). The best description of the experimental data has been obtained using rapidity dependence d$Y$($Y$) (see Fig.6 right). 

Fig.7 demonstrates calculations of dependence of the $\overline{^3\textrm{He}}$ to $^3\textrm{He}$ yield ratios in the central rapidity region on the rapidity and energy in the most central $Cu+Cu$ \cite{He3_ratio_STAR_CuCu} and $Au+Au$ \cite{He3_ratio_STAR} collisions.

\begin{figure}[t]
	\centerline{\includegraphics*[width=1\linewidth]{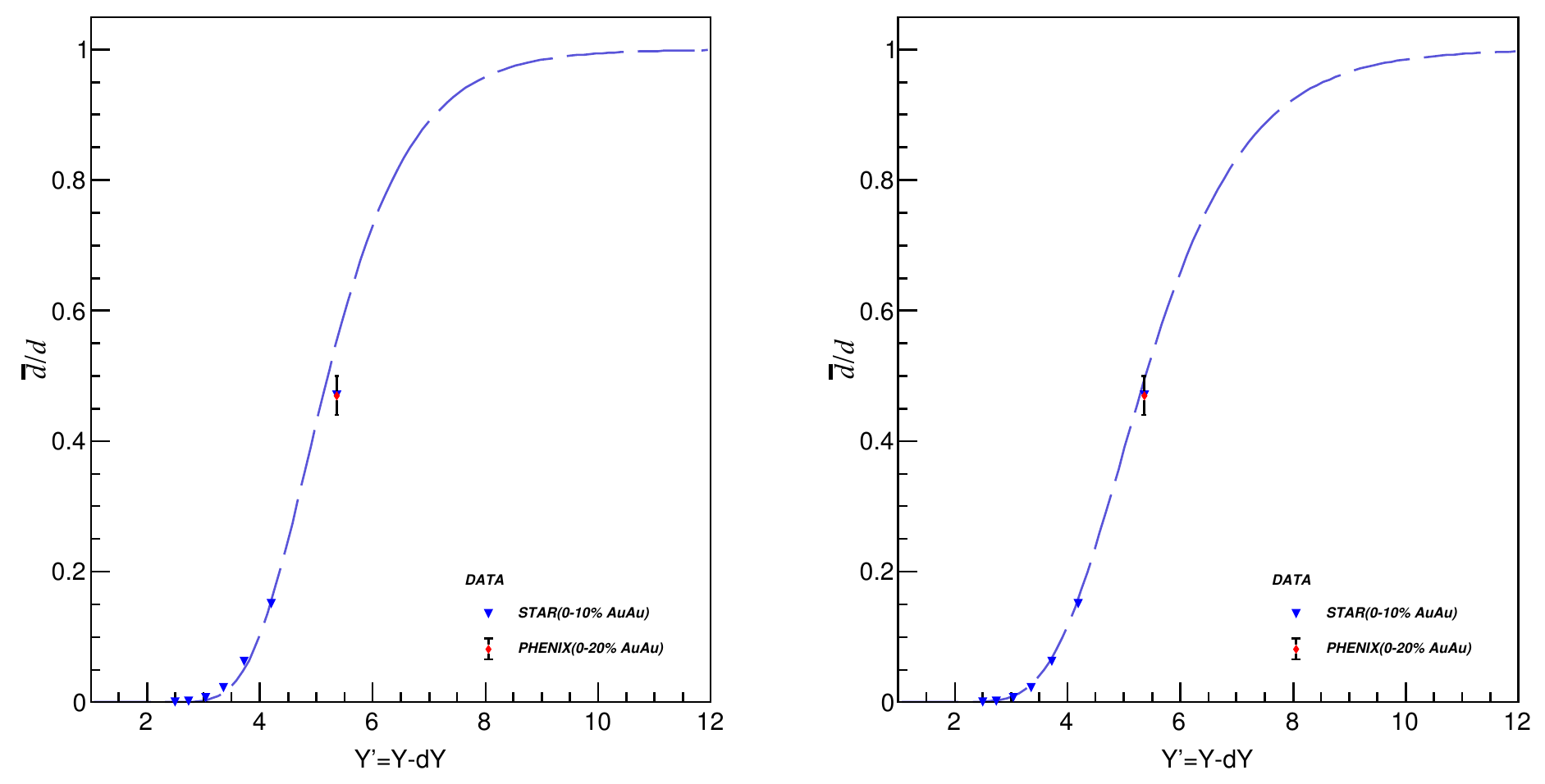}}
	\caption{The calculations of dependence of yield ratios of anti-deuterons to deuterons in the central rapidity region on the rapidity and energy of interacting Au+Au nuclei. On the left side - simple rapidity shift (constant shift), on the right side - using d$Y$($Y$) dependence from Fig.4.}
	\label{fig:Fig.6}
\end{figure}

\begin{figure}[t]
	\centerline{\includegraphics*[width=1\linewidth]{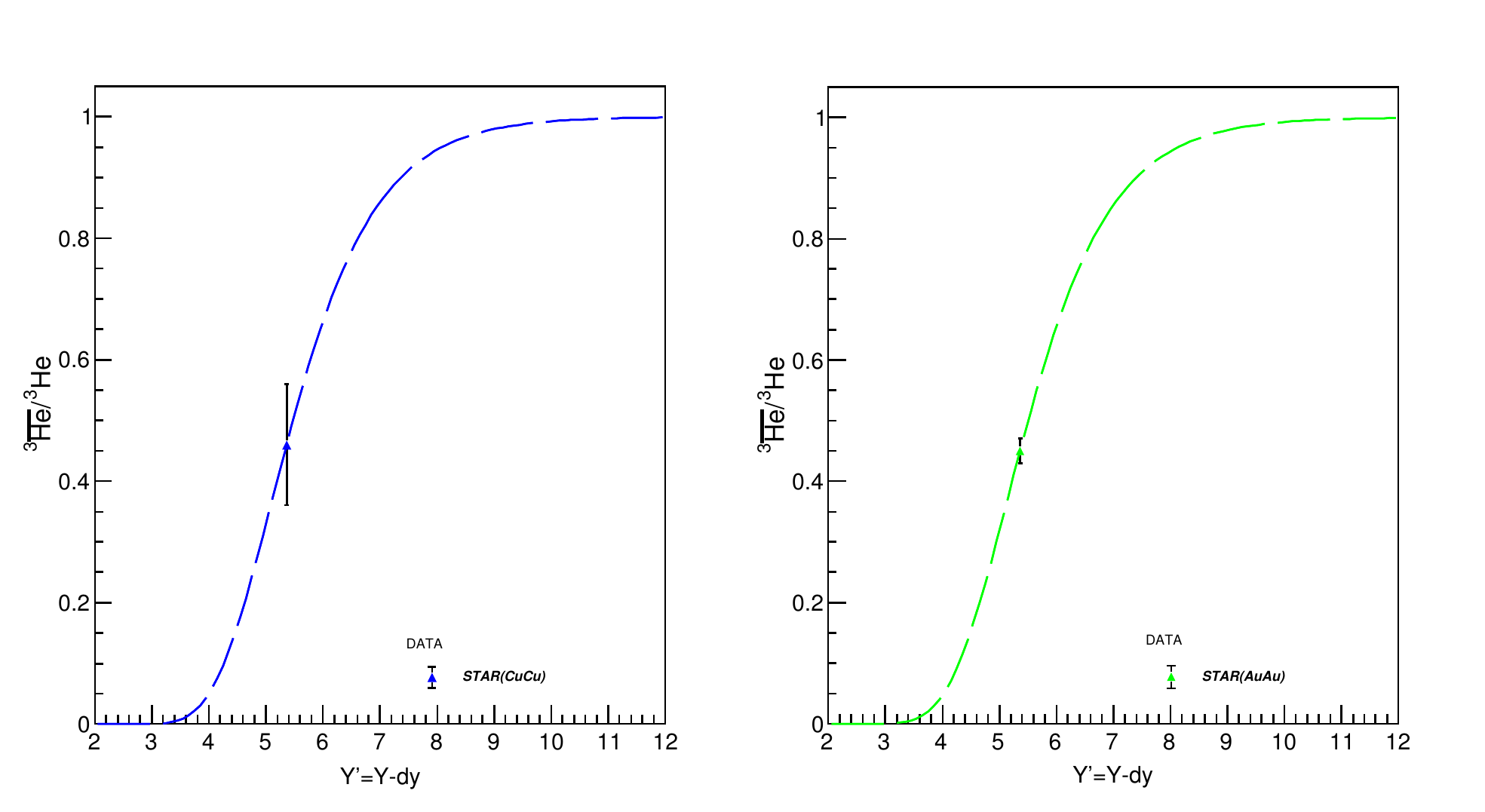}}
	\caption{Dependence of anti-helium-3 to helium-3 yield ratios in the central rapidity region is shown on the left side $Cu+Cu$ collisions, on the right - $Au+Au$ collisions. Simple rapidity shift was used.}
	\label{fig:Fig.7}
\end{figure}

\section*{Conclusion}

In this paper we have successfully applied the self-similarity parameter approach for the fist time to calculate the ratios of anti-baryons to baryons in the mid-rapidity region in a wide range of colliding energies and interacting nuclei. This approach turned out to be very fruitful to describe the ratios of the yield of anti-protons to the yield of protons in $pp$ collisions from $SPS$ to $LHC$ energies. In the case of nuclei-nuclei interactions the rapidity losses must be taken into account while calculating. Taking this fact into account, our approach made it possible to satisfactorily describe the experimental data on the yield ratios of anti-protons to protons, anti-deuterons to deuterons and anti-helium-3 to helium-3 in the central rapidity region in a wide range of colliding nuclei from $S+S$ to $Pb+Pb$. It should be also emphasized that this approach has already been applied successfully to describe inclusive spectra of pions and kaons produced in $pp$ \cite{LMZ:2021} and central $AA$ \cite{ML_2018} collisions as functions of their transverse momentum $p_t$ in the mid-rapidity region. 
 
\begin{acknowledgements}
The authors express their deep gratitude to professors A.G. Litvinenko, G.I. Lykasov, G.L. Melkumov and S.S. Shimansky for the fruitful discussion of the results obtained and also S.V.Chubakova for her help in the design of the manuscript.
\end{acknowledgements}

\end{document}